\documentclass[twocolumn,showpacs,preprintnumbers,amsmath,amssymb,nofootinbib]{revtex4}
\usepackage{graphicx}
\usepackage{dcolumn}
\usepackage{bm}
\def\d{\partial}

\begin{document}

\title{Gravity dual of ${\cal N}\!\!=\!4$ SYM theory with fast moving sources}
\author{Guillaume Beuf}
\email{gbeuf@quark.phy.bnl.gov}
\affiliation{Institut de Physique Th{\'e}orique, CEA/Saclay, 91191 Gif-sur-Yvette cedex,
France \\
URA 2306, unit{\'e} de recherche associ{\'e}e au CNRS}

\begin{abstract}
A family of wave solutions to the full Einstein equations in $AdS_5$ geometry is derived. These background solutions give by duality the response of ${\cal N}\!\!=\!4$ SYM at strong coupling to an arbitrary distribution of fast moving external sources for the energy-momentum tensor operator. We discuss the similarities of these solutions with the Color Glass Condensate effective theory for QCD at high energy.
\end{abstract}


\maketitle


\section{\label{sec:Intro}Introduction}
The AdS/CFT correspondence \cite{Maldacena:1997re,Gubser:1998bc,Witten:1998qj} and other gauge/gravity dualities have brought new insights into strongly coupled gauge theories.
Since the data have provided evidences that the hot and dense matter produced in heavy ion collisions at RHIC should be strongly coupled \cite{Shuryak:2004cy}, AdS/CFT has become an interesting tool in order to explore subjects related to RHIC physics. The quark-gluon plasma (QGP) presumably produced at RHIC is deconfined and at least approximately locally thermalized. In this regime, QCD acquires some similarities with ${\cal N}\!\!=\!4$ SYM at finite temperature, which can justify the use of AdS/CFT. In this context, the hydrodynamical behavior of the long wavelength modes of ${\cal N}\!\!=\!4$ SYM at finite temperature has been understood and related to black brane
physics \cite{Policastro:2002se,Bhattacharyya:2008jc}. This has allowed to calculate the shear viscosity \cite{Policastro:2001yc} and other transport coefficients \cite{Heller:2007qt,Baier:2007ix,Natsuume:2007ty} of a ${\cal N}\!\!=\!4$ SYM plasma. Assuming the symmetries of the Bjorken flow, 
it has been shown that hydrodynamics is the only consistent behavior at late proper time for the strongly coupled plasma \cite{Janik:2005zt,Heller:2009zz,Kinoshita:2008dq}.
The picture of the plasma  emerging from these studies is in qualitative agreement with the one extracted from the data.

Gluon saturation \cite{Gribov:1984tu} dominates the dynamics of the nucleus before the collision, and determines the initial conditions for the medium produced. The correct framework allowing to address that phenomenon from QCD first principles is the Color Glass Condensate effective theory (CGC) (see \cite{Iancu:2002xk,Iancu:2003xm} for reviews), or equivalent formalisms, such as the one developed by Balitsky \cite{Balitsky:2001gj}. The coupling is weak at high enough energy, due to the properties of gluon saturation. Among other successes, the CGC provided a correct prediction for d-Au collisions at RHIC \cite{Kharzeev:2003wz}, where initial condition effects are essential.

One of the main theoretical problems about heavy ion collisions is to understand the thermalization process leading from the weakly coupled out-of-equilibrium Glasma \cite{Lappi:2006fp}, produced by the collision of two sheets of CGC, to the strongly coupled and thermalized QGP.
Let us consider the possibility that thermalization is mainly a strong coupling phenomenon. Then, thermalization should be addressed by building the gravity dual of an expanding medium, starting from very short proper time. In that regime, the physics is determined by the properties of the nucleus before the collision. Hence, it would be useful to construct the gravity dual of a fast moving object mimicking the physics of the CGC.

The obvious problem is that, as a conformal theory, ${\cal N}\!\!=\!4$ SYM admits no states analogous to nucleus or hadrons. There are two main strategies allowing to avoid that difficulty. The first strategy, which is commonly used, is to modify the gravity background dual to ${\cal N}\!\!=\!4$ SYM,
in order to obtain confinement in the IR.
It is however difficult to maintain a gauge/gravity correspondence in that context. Less drastic modifications of the gravity setup such as the introduction of point-like sources in the bulk \cite{Nastase:2004pc,Gubser:2008pc} seem to provide models for a fast moving nucleus in the boundary theory. However, the modification of the boundary field theory corresponding to the addition of sources in the bulk is not well understood. The second possible strategy, developed in the present paper, is the following. We restrict ourselves to the original AdS/CFT correspondence \cite{Maldacena:1997re,Gubser:1998bc,Witten:1998qj}, keeping consistently non-zero sources of ${\cal N}\!\!=\!4$ SYM, and we make use of the results of holographic renormalization \cite{de Haro:2000xn,Bianchi:2001kw}. Suitably chosen distributions of sources generate ${\cal N}\!\!=\!4$ SYM fields whose energy-momentum tensor is a reasonable approximation to the one of a fast moving nucleus in QCD. Such setup is well defined both on the field theory side and on the gravity side of the correspondence, and has some similarities with the weakly coupled CGC.

In section \ref{sec:grav}, we derive a family of background solutions of Einstein equations. In section \ref{sec:holoRenorm}, the boundary field theory counterparts of these solutions are given, relying on the results of holographic renormalization.
In section \ref{sec:CGC}, we discuss the similarities and differences of our solutions with the CGC formalism, and suggest how to model a highly boosted nucleus at strong coupling.

\section{A family of exact solutions of Einstein equations}\label{sec:grav}
Let us look for gravity duals of ${\cal N}\!\!=\!4$ SYM with fast moving matter in the $(+)$ direction on the light-cone. The corresponding metric writes as the $AdS_5$ metric plus an additional part to be found. In the high energy limit, we keep for that new part only the component which is enhanced twice by Lorentz factors with respect to the rest frame of the matter distribution. Hence, generalizing the shockwave solution proposed in \cite{Janik:2005zt}, we consider the Ansatz\footnote{We write with capital letters $I,J,...$ 5-dimensional $AdS_5$-like indices, with Greek letters $\mu, \nu,...$ the 4-dimensional Minkowski-like indices, and with small Latin letters $i,j,...$ indices for the 2-dimensional transverse plane.}  for the metric $G_{IJ}$, in light-cone coordinates,
\begin{equation}
ds^2=\frac{L^2}{z^2} \Big[dz^2 -2dx^+ dx^- +d{\textbf{x}_\bot}^2 + b(x^+,x^-, \textbf{x}_\bot, z) d{x^-}^2\Big]\, ,\label{Ansatz}
\end{equation}
where $\textbf{x}_\bot$ is a vector of the 2-dimensional transverse plane.
We do not modify other supergravity background fields, so that the cosmological constant should remain $\Lambda=-6/L^2$. Hence, one can write the Einstein equations as
\begin{equation}
{\cal R}_{I J}=-\frac{4}{L^2}\ G_{I J}\, .\label{EinsteinEq}
\end{equation}
The $(z,-)$, $(i,-)$ and $(+,-)$ components of the Einstein equations for our Ansatz \eqref{Ansatz} reduce respectively to
\begin{equation}
\d_z \d_+ b=0\, ,\quad \d_i \d_+ b=0\quad \textrm{and}\quad \d_+^2 b=0\, .  \label{EinsteinEqzminus}
\end{equation}
In the function $b(x^+,x^-, \textbf{x}_\bot, z)$, the linear term in $x^+$ allowed by the equations \eqref{EinsteinEqzminus} does not encode interesting physics of the boundary field theory. Hence, we will restrict ourselves, in the following, to functions $b(x^-, \textbf{x}_\bot, z)$ independent of $x^+$.

Now, the non-vanishing Christoffel symbols for our Ansatz are
\begin{eqnarray}
& & \Gamma^z_{--}=-\frac{1}{2}\ \d_z b(x^-, \textbf{x}_\bot, z) +\frac{1}{z}\ b(x^-, \textbf{x}_\bot, z)\nonumber\\
& & \Gamma^i_{--}=-\frac{1}{2}\ \d_i b(x^-, \textbf{x}_\bot, z)\nonumber\\
& & \Gamma^+_{-I}=\Gamma^+_{I-}=-\frac{1}{2}\ \d_I b(x^-, \textbf{x}_\bot, z)\, ,\label{NewAdSChrist}
\end{eqnarray}
in addition to the Christoffel symbols of pure $AdS_5$ geometry, which turn out to stay unmodified\footnote{There is no summation over repeated indices in formula \eqref{AdSChrist}.}
\begin{eqnarray}
& & \Gamma^z_{ij}=\frac{\delta_{ij}}{z}\nonumber\\
& & \Gamma^z_{+-}=\Gamma^z_{-+}=\Gamma^I_{Iz}=\Gamma^I_{zI}=-\frac{1}{z}\, .\label{AdSChrist}
\end{eqnarray}
Calculating the Ricci curvature ${\cal R}_{I J}$, one finds that all the Einstein equations are trivially satisfied, except the one corresponding to the $(-,-)$ component, which gives
\begin{equation}
\d_z^2 b(x^-, \textbf{x}_\bot, z) - \frac{3}{z} \d_z b(x^-, \textbf{x}_\bot, z) + \Delta_\bot b(x^-, \textbf{x}_\bot, z) =0\, .\label{EinsteinEqminusminus}
\end{equation}
In many studies of gauge/gravity duality, one considers small perturbations on the top of a known background, in order to linearize the problem. It should be remarked that here, by contrast, the perturbation $b$ can be large, as the exact Einstein equations turn out to be linear in $b$, thanks to the peculiar structure of the Ansatz \eqref{Ansatz}.
Performing a Fourier transform with respect to the transverse position $\textbf{x}_\bot$, one gets
\begin{equation}
\d_z^2 \tilde{b}(x^-, \textbf{k}_\bot, z) - \frac{3}{z}\ \d_z \tilde{b}(x^-, \textbf{k}_\bot, z) - \textbf{k}_\bot^2\ \tilde{b}(x^-, \textbf{k}_\bot, z) =0\, .\label{EinsteinEqmmFourier}
\end{equation}

For $\textbf{k}_\bot \neq 0$, the generic solutions of equation \eqref{EinsteinEqmmFourier} write
\begin{eqnarray}
\tilde{b}(x^-, \textbf{k}_\bot, z)&=& \tilde{c}_1(x^-, \textbf{k}_\bot)\ \frac{k_\bot^2 z^2}{2}\ K_2(k_\bot z)\nonumber\\
& &+\ \tilde{c}_2(x^-, \textbf{k}_\bot) \ k_\bot^2 z^2\ I_2(k_\bot z)\, ,\label{Modeskdiff0}
\end{eqnarray}
where $I_2$ and $K_2$ are the modified Bessel functions of first and second kind,  $k_\bot = |\textbf{k}_\bot|$, and $\tilde{c}_1$ and $\tilde{c}_2$ are two arbitrary functions.
$z^2 K_2(k_\bot z)$ decays exponentially to zero as $z\rightarrow \infty$ whereas $z^2 I_2(k_\bot z)$ blows up exponentially. Our Poincar\'e coordinates system covers only half of the $AdS_5$ space. For that reason the points with $z\rightarrow \infty$ are in the interior of the manifold. Hence, only $z^2 K_2(k_\bot z)$ is regular in the bulk. Since the function $b$ disappears from the calculation of the square of the Riemann tensor and any other scalar quantities built from the Riemann curvature, one could think that our Ansatz never leads to curvature singularities. However, a similar family of solutions - up to dimension and signature - has been analyzed \cite{Podolsky:1997ik}, and the singularity $z\rightarrow \infty$, if present, has been shown to be a \emph{p. p. curvature singularity}, as defined in \cite{Hawking}. This remark should hold in our case, meaning that the $z^2 I_2(k_\bot z)$ solutions lead to a naked curvature singularity at $z\rightarrow \infty$.

$z^2 I_2(k_\bot z)$ might be acceptable only as the small $z$ behavior of a solution with a particular additional object in the bulk, mapping to a modification of the IR behavior of the boundary field theory. In the present study, we will not discuss further that possibility, and keep only the regular modes with $z^2 K_2(k_\bot z)$.

The solutions of equation \eqref{EinsteinEqmmFourier} for the $\textbf{k}_\bot=0$ modes writes
\begin{equation}
\tilde{b}(x^-, \textbf{k}_\bot=0, z)= \tilde{d}_1(x^-) + z^4 \ \tilde{d}_2(x^-)\, .\label{Modesk0}
\end{equation}
As for the generic case, we will discard the second term, which should lead to a naked curvature singularity at $z\rightarrow\infty$. One should notice that this singular mode is precisely the shockwave solution proposed in \cite{Janik:2005zt} as a model for a highly boosted nucleus, and used in \cite{Grumiller:2008va,Albacete:2008vs} in order to model nucleus-nucleus collisions or deep inelastic scattering on a nucleus. However, one should realize that such a solution which is singular in the bulk does not map to a solution of pure ${\cal N}\!\!=\!4$ SYM in the vacuum.

As $\frac{k_\bot^2 z^2}{2}\ K_2(k_\bot z)\rightarrow 1$ for $z\rightarrow 0$, we merge the two cases, and write our solution for arbitrary $\textbf{k}_\bot$ as
\begin{equation}
\tilde{b}(x^-, \textbf{k}_\bot, z)= \tilde{c}_1(x^-, \textbf{k}_\bot)\ \frac{k_\bot^2 z^2}{2}\ K_2(k_\bot z)\, ,\label{SolModesReg}
\end{equation}
with $\tilde{c}_1(x^-, \textbf{k}_\bot=0)=\tilde{d}_1(x^-)$.
We have a family of regular background solutions (\ref{Ansatz},\ref{SolModesReg}) of Einstein equations.

In that family, the only solutions which are homogeneous in the $\textbf{x}_\bot$ transverse plane are of the type $b(x^-, \textbf{x}_\bot, z)=b(x^-)$. In that case, performing the change of variable $x^+\mapsto \hat{x}^+=x^+ -\int^{x^-} du\, b(u)/2$, one obtains the undeformed $AdS_5$ space, with $\hat{x}^+$ playing the role of $x^+$. Hence, these homogeneous solutions seems not to encode any non-trivial information about ${\cal N}\!\!=\!4$ SYM. We will come back on that issue later on.

Let us now discuss the content of the solutions (\ref{Ansatz},\ref{SolModesReg}) in terms of the dual field theory.

\section{\label{sec:holoRenorm}Dual field theoretical content from holographic renormalization}
The asymptotic behavior near the boundary of background metrics of the type
\begin{equation}
ds^2=\frac{L^2}{z^2} \Big[dz^2  + g_{\mu \nu}(x^{\rho}, z) dx^{\mu}  dx^{\nu}\Big]\, ,\label{FeffGra}
\end{equation}
has been analyzed in \cite{de Haro:2000xn}. When the $z=0$ boundary is 4-dimensional, one finds the expansion
\begin{eqnarray}
g_{\mu \nu}(x^{\rho}, z)&=& g_{(0)\mu \nu}(x^{\rho}) +z^2\  g_{(2)\mu \nu}(x^{\rho}) + z^4\  g_{(4)\mu \nu}(x^{\rho})\nonumber\\
& & +z^4 \log(z^2)\ h_{(4)\mu \nu}(x^{\rho}) + {\cal O}\left(z^6 \log z \right) ,\label{nearBoundExpansion}
\end{eqnarray}
where $g_{(2)\mu \nu}$ and $h_{(4)\mu \nu}$ are completely determined by $g_{(0)\mu \nu}$. The expectation value of the   energy-momentum tensor $\langle T_{\mu \nu}\rangle$ of the dual conformal field theory living on the boundary is calculated according to the usual AdS/CFT prescription \cite{Gubser:1998bc,Witten:1998qj} and with the appropriate holographic renormalization \cite{de Haro:2000xn,Bianchi:2001kw} as
\begin{equation}
\langle T_{\mu \nu}\rangle = \frac{2}{\sqrt{- \det\ g_{(0)}}}\ \frac{\delta}{\delta g_{(0)}^{\mu \nu}} \left(S_{gr, reg}[g_{(0)}] - S_{gr, ct}[g_{(0)}] \right)\, ,\label{HoloTmunuCalculation}
\end{equation}
where $S_{gr, reg}$ is a regularized version of the on-shell gravitational action, and $S_{gr, ct}$ contains the necessary counterterms. Regularization and renormalization are required due to IR divergences near the boundary in the gravity theory which map to UV divergences in the dual field theory. The general result \cite{de Haro:2000xn} of the calculation \eqref{HoloTmunuCalculation} is\footnote{The traces in formula  \eqref{HoloTmunuResult} are taken with respect to the metric $g_{(0)\mu \nu}$.}
\begin{eqnarray}
\langle T_{\mu \nu}\rangle &\!\!=&\!\!\frac{L^3}{4\pi G_N} \Bigg\{\frac{1}{8}\left[\textrm{Tr}(g_{(2)} g_{(0)}^{-1} g_{(2)})- (\textrm{Tr}\ g_{(2)})^2 \right]g_{(0)\mu \nu} \nonumber\\
&\!\! &\!\!\!\!\!\!\!\!\!\!+g_{(4)\mu \nu}\!-\!\frac{1}{2} (g_{(2)}g_{(0)}^{-1} g_{(2)})_{\mu \nu}\!+\!\frac{1}{4} (\textrm{Tr}\ g_{(2)}) g_{(2)\mu \nu}\Bigg\}\label{HoloTmunuResult}
\end{eqnarray}
up to renormalization-scheme dependent terms proportional to $h_{(4)\mu \nu}$.

If one imposes to have the Minkowski metric on the boundary $g_{(0)\mu \nu} = \eta_{\mu \nu}$ once the functional derivation in equation \eqref{HoloTmunuCalculation} is done, one calculates $\langle T_{\mu \nu}\rangle$ in the vacuum. Hence, on the field theory side, $g_{(0) \mu \nu} - \eta_{\mu \nu}$ corresponds to a distribution of sources of $T_{\mu \nu}$. The gravitational background solutions (\ref{Ansatz},\ref{SolModesReg}) are thus dual to the ${\cal N}\!\!=\!4$ SYM theory with non-vanishing external sources $J_{\mu \nu}$ of the energy-momentum tensor, given by
\begin{eqnarray}
J_{\mu \nu}(x^-,\textbf{x}_\bot)&\!\!=&\!\! g_{(0) \mu \nu}(x^-, \textbf{x}_\bot, z=0) - \eta_{\mu \nu}\nonumber\\
&\!\!=&\!\! \delta_{\mu -} \delta_{\nu -}\ b(x^-, \textbf{x}_\bot, z=0)\, .\label{sourcesNeq4}
\end{eqnarray}
That corresponds to a distribution of sources moving to the right at the speed of light, with arbitrary longitudinal and transverse profiles.

In the case of the solutions (\ref{Ansatz},\ref{SolModesReg}), the only non-vanishing component of $g_{(2)\mu \nu}$, $g_{(4)\mu \nu}$  and $h_{(4)\mu \nu}$ is the (-,-) one. Moreover one has $g_{(0)}^{- -}=0$. Therefore, all traces and other nonlinear terms in the metric coefficients present in \eqref{HoloTmunuResult} and in other formulae given in \cite{de Haro:2000xn} vanish, which leads to
\begin{eqnarray}
\langle T_{\mu \nu}\rangle &\!\!=&\!\!\delta_{\mu -} \delta_{\nu -}\ \frac{L^3}{4\pi G_N} \ g_{(4)- -}\label{HoloTmunuResult2}\\
h_{(4)- -}&\!\!=&\!\!-\frac{1}{8}\ \Delta_\bot g_{(2)- -}\label{h4g2pos}\\
g_{(2)- -}&\!\!=&\!\!\frac{1}{4}\ \Delta_\bot g_{(0)- -}\label{g2g0pos}\, .
\end{eqnarray}
Thanks to the expansion\footnote{$\gamma_E\simeq 0.577$ is the Euler-Mascheroni constant.}
\begin{eqnarray}
\frac{k_\bot^2 z^2}{2}\ K_2(k_\bot z)&\!\!=&\!\!1-\frac{k_\bot^2 z^2}{4}+\frac{k_\bot^4 z^4}{32} \Bigg[-2\log \left(\frac{k_\bot z}{2}\right)\nonumber\\
&\!\! &\!\!\; +\frac{3}{2}-2\gamma_E \Bigg]+ {\cal O}\left(z^6 \log z\right)\, ,\label{z2K2expansion}
\end{eqnarray}
one can identify the transverse Fourier transform of the coefficients of the expansion of the metric (\ref{Ansatz},\ref{SolModesReg}) near the boundary, as
\begin{eqnarray}
& &\!\!g_{(0)- -}(x^-, \textbf{k}_\bot)=\tilde{c}_1(x^-, \textbf{k}_\bot)\label{g0imp}\\
& &\!\!g_{(2)- -}(x^-, \textbf{k}_\bot)=-\frac{k_\bot^2}{4}\ \tilde{c}_1(x^-, \textbf{k}_\bot)\label{g2imp}\\
& &\!\!g_{(4)- -}(x^-, \textbf{k}_\bot) +
 \log(z^2)\ h_{(4)- -}(x^-, \textbf{k}_\bot)\nonumber\\
&\!\!=&\!\!\frac{k_\bot^4}{32} \left[-2\log \left(\frac{k_\bot z}{2}\right)+\frac{3}{2}-2\gamma_E \right]\tilde{c}_1(x^-, \textbf{k}_\bot)\label{g4plush4imp}\, .
\end{eqnarray}
From the expressions \eqref{g0imp}, \eqref{g2imp} and \eqref{g4plush4imp}, one checks that the relations \eqref{h4g2pos} and \eqref{g2g0pos} are satisfied, and thus the corresponding generic relations found in \cite{de Haro:2000xn} are also satisfied, as they should.
As $z$ is dimensionful, the left hand side of \eqref{g4plush4imp} is a formal and ambiguous expression. One has to introduce an arbitrary momentum scale $\mu_R$ in order to write $\log \left(k_\bot z/2\right)=\log \left(k_\bot/\mu_R\right)+\log \left(\mu_R z/2\right)$ in the right hand side of \eqref{g4plush4imp}. The ambiguity in the choice of $\mu_R$ reflects the ambiguity in the counterterms used in the formula \eqref{HoloTmunuCalculation}, and thus the choice of the renormalization scale in the boundary field theory. Moreover, it is possible to split arbitrarily the constant terms in the brackets in equation \eqref{g4plush4imp} into a contribution to $g_{(4)- -}$ and a contribution to $\log(z^2)\ h_{(4)- -}$. This last property reflects the freedom in the choice of renormalization scheme in the boundary field theory. Choosing for simplicity to put as many terms as possible in $\log(z^2)\ h_{(4)- -}$, one finds the Fourier transform of the only non vanishing component of $\langle T_{\mu \nu}\rangle$
\begin{eqnarray}
\langle \tilde{T}_{--}\rangle(x^-, \textbf{k}_\bot)&\!\!=&\!\!\frac{L^3 }{4\pi G_N}\ \frac{k_\bot^4}{32}\ \log \left(\frac{\mu_R^2}{k_\bot^2}\right)\ \tilde{c}_1(x^-, \textbf{k}_\bot)\nonumber\\
&\!\!=&\!\!\frac{N_c^2 k_\bot^4}{64\pi^2} \log \left(\frac{\mu_R^2}{k_\bot^2}\right)\tilde{J}_{--}(x^-, \textbf{k}_\bot) \label{HoloTmunuResult3}
\end{eqnarray}
using the AdS/CFT dictionary to express the 5-dimensional Newton constant as $G_N=\frac{\pi L^3}{2 N_c^2}$. One should remark that $\langle {\tilde{T}_{\mu}}^{\mu}\rangle=0$. Hence, the presence of the fast moving external sources does not bring conformal anomalies for ${\cal N}\!\!=\!4$ SYM.

Due to the explicit presence of sources on the gauge theory side, our construction can be understood as an operator deformation of the AdS/CFT correspondence, by the $T_{--}$ operator of ${\cal N}\!\!=\!4$ SYM. Since the metric perturbation $b(x^-, \textbf{x}_\bot, z)$ has a finite limit on the boundary, this deformation is an exactly marginal one. Due to its dimension 4, $T_{\mu \nu}$ is indeed a marginal operator of ${\cal N}\!\!=\!4$ SYM. Hence, in the ultraviolet on the gauge theory side, we do not have the pure ${\cal N}\!\!=\!4$ SYM theory but exactly marginal deformations of it.

Before closing that section, let us come back to the homogeneous solutions $b(x^-, \textbf{x}_\bot, z)=b(x^-)$. From the general results of \cite{de Haro:2000xn}, one finds easily that these solutions correspond in the dual gauge theory to a distribution of sources $J_{--}(x^-)$ which do not induce any non-trivial $\langle T_{--}\rangle(x^-)$, confirming the triviality of these homogeneous solutions. Intuitively, one can understand that from the ${\cal N}\!\!=\!4$ SYM side as follows. After a very strong boost along $x^+$ of a generic field configuration, the dominant components of the ${\cal N}\!\!=\!4$ SYM field strength should be the $F^a_{-i}$ ones, and the dominant component of the $\langle T_{\mu \nu}\rangle$ should be $\langle T_{--}\rangle \propto  (F^a_{-i})^2$. In the case of a homogeneous distribution of sources $J_{--}(x^-)$, there is no preferred direction in the transverse plane, and thus $F^a_{-i}$ has to vanish, and $\langle T_{--}\rangle$ as well. By contrast, if $J_{--}(x^-,\textbf{x}_\bot)$ depend on $\textbf{x}_\bot$, the transverse gradient of $J_{--}$ gives a preferred direction in the transverse plane, allowing a non-vanishing $\langle T_{--}\rangle$.\\

\section{\label{sec:CGC}Discussion: Strongly coupled analog of a Color Glass Condensate}
Our background solutions (\ref{Ansatz},\ref{SolModesReg}) are dual to ${\cal N}\!\!=\!4$ SYM with rightmoving sources on the lightcone. The formula \eqref{HoloTmunuResult3} gives the energy and momentum of the SYM fields induced by the sources, including the potential energy of the fields due to the presence of the sources. This picture is reminiscent of the CGC framework, where a highly boosted nucleus is described by fast moving color sources representing  the long-lived partons and by the quasi-classical gluon field $F^a_{i-}$ they induce. The only non-zero component of $\langle T_{\mu\nu}\rangle$ for this gluon field is also $\langle T_{--}\rangle$. An important difference between the two frameworks is that the CGC is naturally formulated in terms of colored fields whereas AdS/CFT gives access only to colorless objects, which complicates a detailed comparison between the two setups.
Supplemented by a distribution of sources both peaked in the longitudinal direction and localized in a finite region of the transverse plane, the metric (\ref{Ansatz},\ref{SolModesReg}) give a dual model for a fast moving nucleus, which can be easily used to study high energy scattering at strong coupling, including effects of confinement on the geometry of the collision. Most observables usually considered are not sensitive to other, truly dynamical, effects of confinement, which happens only during hadronization, and which are not implemented in the standard AdS/CFT approach.
In principle, one can push further the analogy with the CGC framework by doing the calculation of an observable for a given $J_{--}$, and average the result over $J_{--}$ as done in the CGC, in order to take event-by-event fluctuations of the nucleus wave function into account.

In summary, we have derived a large family of wave solutions (\ref{Ansatz},\ref{SolModesReg}) of the full Einstein equations in $AdS_5$ geometry. Relying on the AdS/CFT correspondence and its holographic renormalization, we have shown that these gravity backgrounds are dual to ${\cal N}\!\!=\!4$ SYM with arbitrary distributions of rightmoving sources on the lightcone. This framework is a strongly coupled analog of the CGC, and provides a laboratory, suitable for the investigation of high energy scattering in ${\cal N}\!\!=\!4$ SYM and thermalization at strong coupling.\\

\paragraph*{{\textbf{Note added:}}} The paper \cite{Avsar:2009xf} appeared on the arXiv a few months after the present manuscript. In \cite{Avsar:2009xf}, the authors discuss deep inelastic scattering on targets given holographically by various shockwave backgrounds, including the ones derived in the present study. The authors of \cite{Avsar:2009xf} discard the $K_2$ solutions we are keeping (see Eq. \eqref{Modeskdiff0}), and consider the $I_2$ solutions we are discarding. However, there is not necessary a contradiction between the two choices, but possibly just two different prescriptions related to two different types of holographic heavy ion collisions.

Indeed, the purpose of \cite{Avsar:2009xf} is to find, using deep inelastic scattering, what kind of shockwave gives the content expected for a nucleus in a confined version of ${\cal N}\!\!=\!4$ SYM at strong coupling. It is shown in \cite{Avsar:2009xf} that the $I_2$ solutions are suitable in that case, provided one modifies the background at large $z$ in order to implement confinement. From the studies of high energy scattering within AdS/CFT, one expects the collision of two such strongly coupled ${\cal N}\!\!=\!4$ SYM nuclei to be quite different from real-world heavy ion collisions \emph{e.g.} at RHIC, which show rather weak stopping properties.

By contrast, the collision of two $K_2$ shockwaves should, at very early proper time, mimic more closely weakly coupled QCD heavy ion collisions, with possibly a Glasma stage. Indeed, the presence of boundary sources allows to impose a high degree of transparency, and an approximate boost-invariance in the central region. However, these $K_2$ shockwaves presumably cannot be obtained from the ultrarelativistic limit of nuclear states of a confined strongly coupled gauge theory.

Moreover, the problem of positive energy condition violation found in \cite{Grumiller:2008va,Albacete:2008vs} for the collision of two $z^4$ shockwaves remains in the case of $I_2$ shockwaves, but should disappear in the case of $K_2$ shockwaves, according to the structure of Einstein equations.

In summary, the $K_2$ solutions maybe more suitable to build an holographic model for real-world QCD heavy ion collisions, although the $I_2$ solutions are more closely related to boosted nuclei in a strongly coupled confined gauge theory.

\vspace{0.4cm}

\begin{acknowledgments}
I acknowledge Iosif Bena, Edmond Iancu, Romuald Janik, Larry McLerran and Robi Peschanski for fruitful discussions and useful comments.
\end{acknowledgments}


\end{document}